\documentclass[aps,preprintnumbers,amsmath,amssymb,a4paper,preprint]{revtex4}
\usepackage{graphicx}
\usepackage{dcolumn}
\usepackage{bm}
\usepackage{color}
\usepackage{ulem}

\begin{document}

\title{Photoelectron circular dichroism with Lissajous-type bichromatic fields: One- vs two-photon ionization of chiral molecules}

\author{Philipp V. Demekhin}\email{demekhin@physik.uni-kassel.de}
\affiliation{Institut f\"ur Physik und CINSaT, Universit\"at Kassel, Heinrich-Plett-Str. 40, 34132 Kassel, Germany}

\begin{abstract}
Angular distribution of photoelectrons released by the ionization of randomly-oriented molecules with two laser fields of carrier frequencies $\omega$ and $2\omega$, which are linearly polarized in two mutually-orthogonal directions, is analyzed in the perturbation limit for the case of one- vs two-photon ionization process. In particular, we focus on the recently predicted [Ph.V. Demekhin {\it et al.}, Phys. Rev. Lett. \textbf{121},  253201 (2018)]  forward-backward asymmetry in the photoelectron emission, which is induced by the interference of two fields and depends on the external relative phase between them. The present theoretical analysis proves a chiral origin of the effect and suggests that a molecule introduces an additional internal relative phase between  two ionizing fields.
\end{abstract}

%\keywords{Calculations and mathematical techniques in atomic and molecular physics,  Photon interactions with molecules, Optical activity and dichroism, Chiral media}

\maketitle

\section{Introduction}

The photoelectron circular dichroism (PECD, \cite{Ritchie}) is a very promising tool for the chiral recognition of molecules in the gas phase  \cite{CDbook,REV1,REV2,REV3}.  The effect consists in the forward-backward asymmetry in the emission of photoelectrons and can be triggered via one-photon  \cite{Expt1,Expt2} or  multiphoton \cite{Lux12,Lehmann13} ionization of chiral molecules. Traditionally, investigation of PECD utilizes circularly (or elliptically \cite{Lux15CPC,Comby16}) polarized light. The observed effect emerges in the electric-dipole approximation \cite{Ritchie} as an incomplete compensation of the amplitudes for the emission of partial photoelectron waves with positive and negative projections $m$ of the carried angular momentum $\ell$. Here, a typical forward-backward asymmetry can reach about 10\% of the total photoionization signal \cite{CDbook,REV1,REV2,REV3}.

\begin{figure}
\includegraphics[scale=0.22]{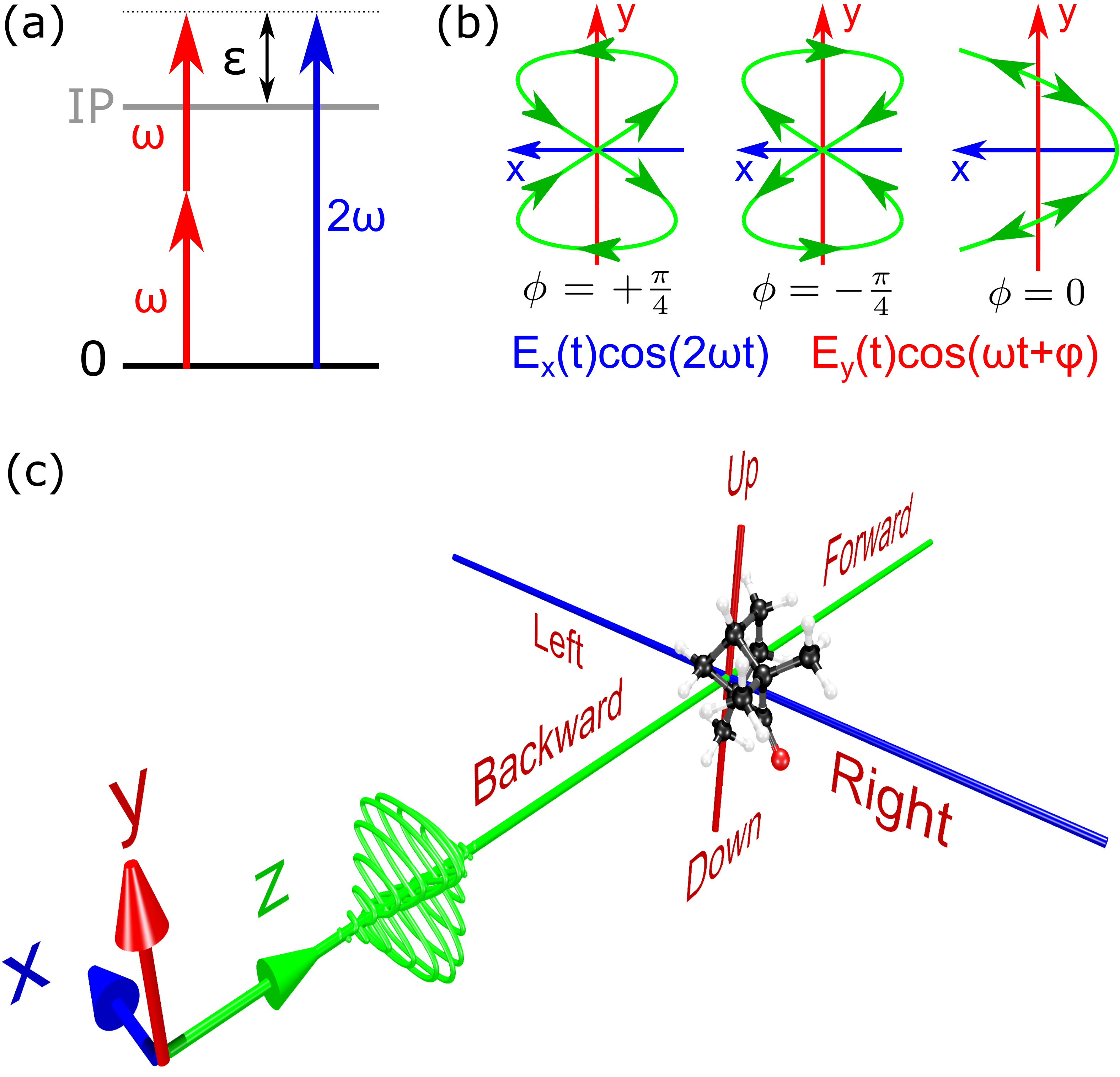}
\caption{{Panel (a): Energy scheme of the one- vs two-photon ionization of a chiral molecule from its ground electronic state (set to the energy origin) in the electronic continuum state of energy $\varepsilon=2\omega-IP$, where $IP$ stays for the ionization potential. Panel (b): Different configurations of the total electric field (\ref{eq:field1}) for different values of the relative phase $\phi$. Directions of rotation along the field trajectory are indicated by arrows. Panel (c): The system of coordinates and notations utilized in the present work. The field propagates along $z$-axis, while $x$- and $y$-axes are defined according to Eq.~(\ref{eq:field1}).}}\label{fig}
\end{figure}

Recently \cite{PRLw2w}, it was suggested that PECD can also be observed by a Lissajous-type electric field configuration consisting of carrier frequencies $\omega$ and $2\omega$  linearly polarized in two mutually-orthogonal directions:
\begin{equation}
\mathcal{\vec{E}}(t)= \hat{e}_x \mathcal{E}_x  \cos(2\omega t) + \hat{e}_y \mathcal{E}_y  \cos(\omega t +\phi).
\label{eq:field1}
\end{equation}
Depending on the relative phase $\phi$ between two fields, the resulting electric field vector mimics rotational motions in different directions in the dipole $xy$-plane, which is perpendicular to the propagation of the light (along $z$-axis). For instance, for $\phi=\pm \frac{\pi}{4}$, the electric field (\ref{eq:field1}) forms a `butterfly' that is oriented along the $y$-axis [{see Fig.~\ref{fig}(b)}]. Such a field possesses  in the upper and lower hemispheres two different rotational directions and induces thereby two opposite forward-backward asymmetries. This subcycle chiral asymmetry {emerges due} to the interference between two fields.

Numerical calculations, performed in Ref.~\cite{PRLw2w} with the time-dependent single center method \cite{TDSC1,TDSC2} for the one- vs two-photon ionization of a model methane-like chiral system, demonstrate a sizable PECD which depends on the handedness of a chiral molecule, rotational direction of the field, and the relative  phase $\phi$ between two fields. In the present work, we investigate the chiral asymmetry  predicted in Ref.~\cite{PRLw2w} for the one- vs two-photon ionization process in details. In particular, we derive and analyze the emerging PECD signal in the weak-field limit in terms of the partial photoionization amplitudes. {Energy scheme of the considered process is depicted in  Fig.~\ref{fig}(a).}

\section{Theoretical analysis}

Below, we utilize the system of coordinates and notations introduced in {Figs.~\ref{fig}(b,c)}. To assist tracing the present derivation, the most important used relations, {adjusted from the textbook Ref.~\cite{Varshalovich} to the present notations,} are collected in the appendix.

\subsection{Basic equations}

In the weak-field limit, the electric field (\ref{eq:field1}) can be split into two parts, which in the  rotating wave approximation are responsible for the absorption and emission of photons:
\begin{equation}
\mathcal{\vec{E}}= \frac{1}{2} \left(\hat{e}_x \mathcal{E}_x  e^{-2i\omega t} + \hat{e}_y \mathcal{E}_y e^{-i\omega t} e^{-i\phi} \right) +cc(\mathrm{emission}).
\end{equation}
In order to unambiguously introduce a unique direction of the propagation of both fields, we make use of {spherical basis} (\ref{eq:cyclicA}):
\begin{equation}
\label{eq:cyclic}
\hat{e}_x=\frac{1}{\sqrt 2}\left(-\hat{e}_+ + \hat{e}_- \right), ~~~~\hat{e}_y=\frac{i}{\sqrt 2}\left(\hat{e}_+ + \hat{e}_- \right).
\end{equation}
Here, `$\pm$' correspond to the circularly polarized light with positive or negative helicity, both propagating along the laboratory $z$-axis. In the dipole length gauge, the respective light-matter interaction reads
\begin{equation}
\label{eq:operqtor}
\vec{r}\cdot\mathcal{\vec{E}}=  \frac{\mathcal{E}_x }{2\sqrt 2}\left(- \mathbf{d}_{+1}  + \mathbf{d}_{-1}  \right) + \frac{i\mathcal{E}_y }{2\sqrt 2}\left( \mathbf{d}_{+1}  + \mathbf{d}_{-1}  \right)e^{-i\phi} ,
\end{equation}
with the electric dipole  operator $ \mathbf{d}_{k}=r \sqrt{\frac{4\pi}{3}}Y_{1k}$. Here, $k=0$ and $k=\pm1$ stand for the linear and circular polarizations, respectively.

The wave function of a photoelectron in the continuum spectrum of energy $\varepsilon={p^2}/{2}$ is described by the incoming-wave-normalized \cite{Starace} superposition of spherical waves with given angular momentum quantum numbers $\ell$ and $m^\prime$ (known as the partial electron waves \cite{Cherepkov81}):
\begin{equation}
\label{eq:photelectron}
\Psi_\mathbf{p}^-(\mathbf{r^\prime}) = \sum_{\ell m^\prime} (i)^{\ell}  R_{\varepsilon \ell m^\prime}(r^\prime) \, Y_{\ell m^\prime}(\theta^\prime,\varphi^\prime) \,Y^\ast_{\ell m ^\prime}(\theta_p^\prime,\varphi_p^\prime).
\end{equation}
Here, prime refers to the  quantum numbers and coordinates as defined in the frame of the molecule. Note also that the normalization coefficient and  the phases of the partial waves are incorporated in the respective radial parts $R_{\varepsilon \ell m^\prime}$, for brevity.

In the frame of molecule, the electric dipole transition matrix elements for the emission of the partial wave $\varepsilon \ell m^\prime$ by the absorption of one photon of energy $2\omega$ and polarization $k^\prime$ reads:
\begin{equation}
\label{eq:onephotonME}
d_{\ell m ^\prime k^\prime}=\langle R_{\varepsilon \ell m^\prime}Y_{\ell m^\prime} \vert \mathbf{d}_{k^\prime} \vert \Phi_0 \rangle _{_{\mathbf{r}^\prime}}.
\end{equation}
{Here and below, the subscript $\mathbf{r}^\prime$ indicates an integration over spatial coordinates in  the frame of molecule.} Since absorption of one photon of energy $\omega$ is not sufficient to ionize the initial molecular orbital $\Phi_0$, the respective two-photon transition matrix element reads:
\begin{equation}
\label{eq:twophotonME}
{t}_{\ell m ^\prime k_1^\prime k_2^\prime}=\sum_I \frac{\langle R_{\varepsilon \ell m^\prime} Y_{\ell m^\prime} \vert \mathbf{d}_{k_2^\prime} \vert \Phi_I  \rangle_{_{\mathbf{r}^\prime}} \langle \Phi_I  \vert \mathbf{d}_{k_1^\prime} \vert \Phi_0 \rangle_{_{\mathbf{r}^\prime}}}{E_I-E_0-\omega}.
\end{equation}
Here, summation must be performed over all virtual intermediate electronic states $\Phi_I$ in the discrete and continuum electron spectrum.

In order to proceed, we transform all electric dipole operators entering Eqs.~(\ref{eq:onephotonME}) and (\ref{eq:twophotonME}) from the laboratory to  the molecular frame and the emitted partial waves from Eq.~(\ref{eq:photelectron}) in the opposite way. The transformations (\ref{eq:photonA}) and (\ref{eq:electronA}) are performed  with the help of the Wigner's rotation matrices $\mathcal{D}^\ell_{i,j}$ for given Euler orientation angles $\left\{\alpha,\beta,\gamma\right\}$. Making use of the introduced basic equations, the total amplitude of the considered process can be written as a coherent superposition of the amplitudes for the one-photon  (\ref{eq:onephotonME})  and two-photon (\ref{eq:twophotonME}) ionizations:
\begin{multline}
\label{eq:TOTAMPL}
\mathbb{D}_{T}=\sum_{\ell_1 m_1^\prime} (-i)^{\ell_1}  \sum_{m_1}  \mathcal{D}^{\ell_1\ast}_{m_1^\prime,m_1} Y_{\ell_1 m_1 }(\hat{p})\left[ \frac{\mathcal{E}_x }{2\sqrt 2} \sum_{k^\prime}\left( - \mathcal{D}^1_{k^\prime,+1}  +  \mathcal{D}^1_{k^\prime,-1}  \right) d_{\ell_1 m_1 ^\prime k^\prime}\right. \\ \left.-e^{-2i\phi} \frac{\mathcal{E}^2_y }{ 8} \sum_{k_1^\prime k_2^\prime}\left( \mathcal{D}^1_{k_1^\prime,+1} \mathcal{D}^1_{k_2^\prime,+1} + \mathcal{D}^1_{k_1^\prime,+1} \mathcal{D}^1_{k_2^\prime,-1} + \mathcal{D}^1_{k_1^\prime,-1} \mathcal{D}^1_{k_2^\prime,+1}+\mathcal{D}^1_{k_1^\prime,-1} \mathcal{D}^1_{k_2^\prime,-1}   \right) {t}_{\ell_1 m_1 ^\prime k_1^\prime k_2^\prime}\right].
\end{multline}
As one can see from Eq.~(\ref{eq:TOTAMPL}), the interference between the one- and two-photon routes (the sum of respective terms in the square brackets) can be controlled by the relative phase $\phi$ between two fields (i.e., by the factor $e^{-2i\phi}$).

\subsection{Analysis of the asymmetry}

We first notice that the product $\mathbb{D}_{T}\mathbb{D}^\ast_{T}$ contains the product of two spherical functions which define the direction of the emission of the photoelectron $\hat{p}=(\theta_p,\varphi_p)$ in the laboratory frame. The latter product can be reduced to the sum over spherical functions via Eq.~(\ref{eq:directionA}). As a consequence, the differential cross section for the emission of photoelectrons can be expanded as
\begin{equation}
\label{eq:dcs}
\frac{d\sigma}{d\Omega_p} =2\pi \vert \mathbb{D}_{T}\vert^2=\sum_{LM} B_{LM}Y^\ast_{LM }(\hat{p}).
\end{equation}
Expansion (\ref{eq:dcs}) has three different contributions. The product of the one-photon ionization amplitudes (\ref{eq:onephotonME}) includes terms with  $L\leq2$, and the product of the two-photon ionization amplitudes (\ref{eq:twophotonME}) is restricted by $L\leq 4$ {\cite{AD1,AD2}}. The cross-products of the one- and two-photon ionization amplitudes (i.e., the interference terms between two fields which are responsible for the effect  predicted in Ref.~\cite{PRLw2w}) are restricted by $L\leq3$. In addition, according to symmetry considerations, a combined {\it forward-backward} (with respect to the $\pm z$ directions) and {\it up-down} (with respect to the $\pm y$ directions) asymmetry is described by the expansion terms $B_{LM}$ with even values of $L$ and odd values of $M$. Below, this combined asymmetry is referred to as $FBUD$ asymmetry.

As justified above, in the case of one- vs two-photon ionization process, the $FBUD$ asymmetry dictated by the interference between two fields is given by:
\begin{equation}
\label{eq:FBUD1}
FBUD(\hat{p})=B_{2,+1}Y^\ast_{2,+1 }(\hat{p})+B_{2,-1}Y^\ast_{2,-1 }(\hat{p}),
\end{equation}
{which are the only terms with even $L$ and odd $M$ values from the expansion limited by $L\leq3$}. In order to keep this observable real, the condition $B_{2,-1}=-B^\ast_{2,+1}$ should be fulfilled. Using now the explicit expressions (\ref{eq:Y2mp1A}) for the spherical functions $Y^\ast_{2,\pm1 }$, we obtain
\begin{equation}
\label{eq:FBUD2}
FBUD(\theta_p,\varphi_p)= -\sqrt\frac{15}{2\pi}\cos\theta_p\sin\theta_p \left[\mathrm{Re}(B_{2,+1})\cos\varphi_p + \mathrm{Im}(B_{2,+1})\sin\varphi_p\right].
\end{equation}
Further symmetry considerations suggest that a  combined {\it forward-backward} and {\it left-right} (with respect to the $\pm x$ directions) asymmetry is absent for all relative phases $\phi$. {This is because electric field (\ref{eq:field1}) always possesses two opposite rotational directions for equal periods of time in each of the left and right hemispheres. Thus, the term with $\cos\varphi_p$ in Eq.~(\ref{eq:FBUD2}) must vanish.} This holds only if $B_{2,+1}$ is purely imaginary, i.e., $\mathrm{Re}(B_{2,+1})=0$. Taking into account that there are two complex conjugate cross-terms in the $\mathbb{D}_{T}\mathbb{D}^\ast_{T}$ product, this expansion coefficient can be represented as:
\begin{equation}
\label{eq:b21}
B_{2,+1}=\mathcal{A}e^{-2i\phi}-\mathcal{A}^\ast e^{2i\phi}= -2i{A}\sin(2\phi-\delta)  ,
\end{equation}
with the complex coefficient $\mathcal{A}={A}e^{i\delta}$. We finally arrive at the following equation for the $FBUD$ asymmetry:
\begin{equation}
\label{eq:FBUD3}
FBUD(\theta_p,\varphi_p)=   A \cos\theta_p\sin\theta_p \sin\varphi_p \sin(2\phi-\delta).
\end{equation}
Note, that the remaining prefactor $\sqrt{30/\pi}$ is now explicitly included in the complex coefficient $\mathcal{A}$ from the  expression~(\ref{eq:b21}) for $B_{2,+1}$.

\subsection{Magnitude of the asymmetry }

In order to compute the coefficient $\mathcal{A}$, we set $L=2$ and $M=+1$ in the expansion (\ref{eq:dcs}) and collect all terms in front of the phase $e^{-2i\phi}$:
\begin{widetext}
\begin{multline}
\label{eq:AAA1}
\mathcal{A }=-2\pi\sqrt\frac{30}{\pi} \sum_{\ell_1 m_1^\prime m_1}(-i)^{\ell_1}  \mathcal{D}^{\ell_1\ast}_{m_1^\prime,m_1}  \frac{\mathcal{E}^2_y }{ 8} \sum_{k_1^\prime k_2^\prime}\left( \mathcal{D}^1_{k_1^\prime,+1} \mathcal{D}^1_{k_2^\prime,+1} + \mathcal{D}^1_{k_1^\prime,+1} \mathcal{D}^1_{k_2^\prime,-1} + \mathcal{D}^1_{k_1^\prime,-1} \mathcal{D}^1_{k_2^\prime,+1}\right. \\ \left. + \mathcal{D}^1_{k_1^\prime,-1} \mathcal{D}^1_{k_2^\prime,-1}   \right) {t}_{\ell_1 m_1 ^\prime k_1^\prime k_2^\prime}  \sum_{\ell_2 m_2^\prime m_2} (i)^{\ell_2}    \mathcal{D}^{\ell_2}_{m_2^\prime,m_2} \frac{\mathcal{E}_x }{2\sqrt 2} \sum_{k^{\prime\prime}}\left( - \mathcal{D}^{1\ast}_{k^{\prime\prime},+1}  + \mathcal{D}^{1\ast}_{k^{\prime\prime},-1} \right) d^\ast_{\ell_2 m_2 ^\prime k^{\prime\prime}}   \\ \times (-1)^{m_2}\sqrt\frac{(\ell_1)(\ell_2)5}{4\pi}\left( \begin{array}{ccc} \ell_1 & \ell_2 &2 \\0 & 0& 0\end{array}  \right)\left( \begin{array}{ccc} \ell_1 & \ell_2 &2 \\m_1 & -m_2& +1\end{array}  \right).
\end{multline}
By reducing the products of the rotation matrices via Eqs.~(\ref{eq:DDD1A}) and (\ref{eq:DDD2A}) and performing a summation over the indices $m_1$ and $m_2$  via Eq.~(\ref{eq:DDD3A}),  we simplify Eq.~(\ref{eq:AAA1}) to the following form (note that, according to Eq.~(\ref{eq:DDD3A}), $J=2$ and $M_J=1$):
\begin{multline}
\label{eq:AAA2}
\mathcal{A }=\frac{5\sqrt{3}}{16} \mathcal{E}_x\mathcal{E}^2_y   \sum_{\ell_1 m_1^\prime}\sum_{\ell_2 m_2^\prime } \sum_{k_1^\prime k_2^\prime k^{\prime\prime}} \sum_{M_J^\prime} \sqrt{(\ell_1)(\ell_2)} \, (i)^{\ell_1+\ell_2} \left( \begin{array}{ccc} \ell_1 & \ell_2 &2 \\0 & 0& 0\end{array}  \right) \left( \begin{array}{ccc} \ell_1 & \ell_2 &2 \\-m_1^\prime & m_2^\prime& -M_J^\prime\end{array}  \right) \mathcal{D}^{2}_{M_J^\prime,+1}  \\ \times (-1)^{k^{\prime\prime}}\left[  -\mathcal{D}^{1}_{-k^{\prime\prime},-1} + \mathcal{D}^{1}_{-k^{\prime\prime},+1} \right] d^\ast_{\ell_2 m_2 ^\prime k^{\prime\prime}} \sum_{TM_TM^\prime_T}    (-1)^{\ell_2+m_1^\prime-M_J^\prime+M_T-M^\prime_T}  (T) \left( \begin{array}{ccc} 1 & 1 &T \\ k_1^\prime& k_2^\prime& -M^\prime_T\end{array}  \right) \mathcal{D}^{T}_{M^\prime_T,M_T}  \\ \times \left[ \left( \begin{array}{ccc} 1 & 1 &T \\ +1 & +1  & -M_T\end{array}  \right) + \left( \begin{array}{ccc} 1 & 1 &T \\ +1 & -1  & -M_T\end{array}  \right) +\left( \begin{array}{ccc} 1 & 1 &T \\ -1 & +1  & -M_T\end{array}  \right) +\left( \begin{array}{ccc} 1 & 1 &T \\ -1 & -1  & -M_T\end{array}  \right) \right] {t}_{\ell_1 m_1 ^\prime k_1^\prime k_2^\prime}.
\end{multline}

Finally, we average the product of the three remaining rotation matrices over all molecular orientations via Eq.~(\ref{eq:3rotmatA}). Equations~(\ref{eq:AAA2}) and (\ref{eq:3rotmatA}) suggest that only the values $T=2$ and $M_T=0,-2$ are allowed, which yields:
\begin{multline}
\label{eq:AAA3}
\mathcal{A }=\frac{25\sqrt{3}}{16} \mathcal{E}_x\mathcal{E}^2_y   \sum_{\ell_1 m_1^\prime}\sum_{\ell_2 m_2^\prime } \sum_{M_J^\prime M^\prime_T} \sqrt{(\ell_1)(\ell_2)}\,(i)^{\ell_1+\ell_2}  (-1)^{\ell_2+m_1^\prime}\left( \begin{array}{ccc} \ell_1 & \ell_2 &2 \\0 & 0& 0\end{array}  \right) \left( \begin{array}{ccc} \ell_1 & \ell_2 &2 \\-m_1^\prime & m_2^\prime& -M_J^\prime\end{array}  \right)  \\ \times
 \left[ -\left( \begin{array}{ccc} 2& 1&2\\ +1  & -1&0\end{array}  \right) \left\{ \left( \begin{array}{ccc} 1 & 1 &2 \\ +1 & -1  & 0\end{array}  \right) +\left( \begin{array}{ccc} 1 & 1 &2 \\ -1 & +1  & 0\end{array}  \right)\right\}
+\left( \begin{array}{ccc} 2& 1&2\\ +1  & +1&-2\end{array}  \right) \left( \begin{array}{ccc} 1 & 1 &2 \\ -1 & -1  & 2\end{array}  \right) \right] \\ \times\sum_{k_1^\prime k_2^\prime k^{\prime\prime}} \left( \begin{array}{ccc} 1 & 1 &2 \\ k_1^\prime& k_2^\prime& -M^\prime_T\end{array}  \right)\left( \begin{array}{ccc} 2& 1&2\\ M_J^\prime  & -k^{\prime\prime}  & M^\prime_T \end{array}  \right)d^\ast_{\ell_2 m_2 ^\prime k^{\prime\prime}}  {t}_{\ell_1 m_1 ^\prime k_1^\prime k_2^\prime}.
\end{multline}
By making use of the following relations $M^\prime_T=k_1^\prime+ k_2^\prime $, $M_J^\prime=k^{\prime\prime} - M^\prime_T =k^{\prime\prime} -k_1^\prime- k_2^\prime$, and $ M_J^\prime= m_2^\prime  -m_1^\prime$, the summations over indices $M_J^\prime$ and $M^\prime_T$ in Eq.~(\ref{eq:AAA3}) can be omitted. We now replace the second line of this equation by its explicit value of $\frac{2}{5\sqrt3}$ and arrive at the following final expression for the complex coefficient  $\mathcal{A }$:
\begin{multline}
\label{eq:AAA4}
\mathcal{A }=\frac{5}{8} \mathcal{E}_x\mathcal{E}^2_y   \sum_{\ell_1\ell_2 } \sqrt{(\ell_1)(\ell_2)}\, (i)^{\ell_1+\ell_2}  (-1)^{\ell_2^\prime} \left( \begin{array}{ccc} \ell_1 & \ell_2 &2 \\0 & 0& 0\end{array}  \right)\sum_{ m_1^\prime m_2^\prime } \sum_{k_1^\prime k_2^\prime k^{\prime\prime}} (-1)^{m_1}   d^\ast_{\ell_2 m_2 ^\prime k^{\prime\prime}}  {t}_{\ell_1 m_1 ^\prime k_1^\prime k_2^\prime}   \\  \times\left( \begin{array}{ccc} \ell_1 & \ell_2 &2 \\-m_1^\prime & m_2^\prime&  ( k_1^\prime+k_2^\prime- k^{\prime\prime})\end{array}  \right)   \left( \begin{array}{ccc} 2& 1&2\\   (m_2^\prime  -m_1^\prime ) & -k^{\prime\prime}  & ( k_1^\prime+k_2^\prime) \end{array}  \right)\left( \begin{array}{ccc} 1 & 1 &2 \\ k_1^\prime& k_2^\prime& -(k_1^\prime+k_2^\prime)\end{array}  \right).
\end{multline}
\end{widetext}

\section{Discussion and outlook}

We first prove the chiral origin of the $FBUD$ asymmetry. For this purpose, we notice that only the partial waves of the same parity can contribute to the asymmetry, which is different from the case of a traditional PECD \cite{Ritchie}. This fact follows from the very fist 3j-symbol in Eq.~(\ref{eq:AAA4}), which suggests that $\ell_1+\ell_2$ must be even. We further notice that the second  3j-symbol in the second line of Eq.~(\ref{eq:AAA4}) changes its sign according to
\begin{equation}
\label{eq:asym3j}
\left( \begin{array}{ccc} 2& 1&2\\  (m_2^\prime  -m_1^\prime  ) & -k^{\prime\prime}  & ( k_1^\prime+k_2^\prime) \end{array}  \right) = -\left( \begin{array}{ccc} 2& 1&2\\  -(m_2^\prime  -m_1^\prime )  &  k^{\prime\prime}  & - (k_1^\prime+k_2^\prime) \end{array}  \right).
\end{equation}
Simultaneously, the remaining two 3j-symbols in the second line of Eq.~(\ref{eq:AAA4}) are  symmetric with respect to the interchange of signs in the lower rows (since  sum of upper indices is always even). Therefore, when performing summations  over the complete range of indices $\left\{m_1^\prime,m_2^\prime,k^{\prime\prime}, k_1^\prime,k_2^\prime\right\}$, the respective contributions from the positive and negative angular momentum projections and light helicities will cancel out. Such  a complete cancellation would not occur if the respective electric dipole transition amplitudes were inequivalent:
\begin{subequations}
\begin{equation}
\label{eq:chiral1}
d_{\ell_2 m_2 ^\prime k^{\prime\prime}} \ne  d_{\ell_2, -m_2 ^\prime, -k^{\prime\prime}},
\end{equation}
\begin{equation}
\label{eq:chiral2}
{t}_{\ell_1 m_1 ^\prime k_1^\prime k_2^\prime} \ne {t}_{\ell_1, -m_1 ^\prime, -k_1^\prime, -k_2^\prime},
\end{equation}
\end{subequations}
which is the case for chiral molecules \cite{Ritchie}.

Numerical calculations, performed in Ref.~\cite{PRLw2w} for the one- vs two-photon ionization of a model chiral system, showed that the $FBUD$ asymmetry is maximal for the relative phases $\phi=\pm \frac{\pi}{4}$ [`butterfly' form {in Fig.~\ref{fig}(b)}], and it vanishes for  $\phi=0$ and $\phi=\pm\frac{\pi}{2}$ [`horseshoe' form  {in Fig.~\ref{fig}(b)}]. This result is in agreement with Eq.~(\ref{eq:FBUD3}), if the argument $\delta$ of the complex coefficient $\mathcal{A}$ vanishes. In general, the chiral asymmetry (\ref{eq:FBUD3}) is proportional to $\sin(2\phi-\delta)$, and for particular values of $\delta$ it maximizes and minimizes at  different relative phases $\phi$. The argument $\delta$  of the complex coefficient $\mathcal{A}$ [given by Eq.~(\ref{eq:AAA4})] depends on the transition amplitudes for the one- and two-photon ionization, which are inherent electronic properties of a molecule at a given  frequency $\omega$. It, therefore, can be considered as an internal phase, which is introduced to the fields by a molecule in additional to the external phase $\phi$. In the other words, for $\delta\ne 0$ a molecule `sees' a field configuration which is different from that prepared by the experimental setup. Such an internal molecular phase {\cite{MolPhase}} is commonly utilized in different coherent control schemes \cite{CoCo1,CoCo2,Goetz19}. It is especially relevant for resonance-enhanced multiphoton ionization (REMPI) schemes, where different intermediate electronic states introduce additional phases to the interference between a manifold of multiphoton ionization pathways.

As a final point, we discuss the optical-regime photoionization with bichromatic fields (\ref{eq:field1}), where higher-order multiphoton processes need to be involved. In this case, a manifold of interference terms between different multiphoton ionization pathways will contribute to the chiral asymmetry by several expansion terms $B_{LM}$ with even values of $L$ and odd values of $M$. The respective combined forward-backward and up-down asymmetry will be angularly-structured by the higher order $(\cos\theta_p)^{2n+1}$ and $\sin([2n+1]\varphi_p)$ terms. Moreover, the contributions from different multiphoton pathways will possess different dependencies on the external phase. As photoionization in the optical regime likely involves intermediate resonances, a molecule will introduce different internal phases to those pathways. The respective contributions, which  scale as $\sin([2n]\phi-\delta_n)$, will minimize and maximize at different external phases.

\begin{acknowledgements}
The author acknowledges  T. Baumert and O. Smirnova for many valuable discussions. This work was funded  by the Deutsche Forschungsgemeinschaft (DFG, German Research Foundation) -- Projektnummer 328961117 -- SFB 1319 ELCH (subproject C1).
\end{acknowledgements}

\appendix*
\section{Useful relations}
\label{sec:app}

Definition of {spherical basis}:
\begin{equation}
\label{eq:cyclicA}
\hat{e}_\pm = \mp\frac{1}{\sqrt 2}\left(\hat{e}_x \pm i \hat{e}_y \right).
\end{equation}

Transformation of the electric dipole operator from the laboratory to molecular frame:
\begin{equation}
\label{eq:photonA}
\mathbf{d}_k  =  \sum_{k^\prime}  \mathcal{D}^1_{k^\prime,k}(\alpha,\beta,\gamma)\, \mathbf{d}_{k^\prime} .
\end{equation}

Transformation of the partial emission waves from the molecular to laboratory frame:
\begin{equation}
\label{eq:electronA}
Y_{\ell m ^\prime}(\hat{p}^\prime)  =  \sum_{m}  \mathcal{D}^{\ell\ast}_{m^\prime,m}(\alpha,\beta,\gamma)\,  Y_{\ell m }(\hat{p}) .
\end{equation}

Reduction of the product of two spherical functions:
\begin{multline}
\label{eq:directionA}
Y_{\ell_1 m_1 }(\hat{p})Y^\ast_{\ell_2 m_2 }(\hat{p})=\\(-1)^{m_2}\sum_{LM}\sqrt\frac{(\ell_1)(\ell_2)(L)}{4\pi}\left( \begin{array}{ccc} \ell_1 & \ell_2 &L \\0 & 0& 0\end{array}  \right) \left( \begin{array}{ccc} \ell_1 & \ell_2 &L \\m_1 & -m_2& M\end{array}  \right)Y^\ast_{LM }(\hat{p}),
\end{multline}
where $(\ell)=2\ell+1$ for brevity.

Explicit expressions for spherical functions:
\begin{equation}
\label{eq:Y2mp1A}
 Y^\ast_{2,\pm1 }(\hat{p})=\mp\frac{1}{2}\sqrt\frac{15}{2\pi}\cos\theta_p\sin\theta_p e^{\mp i\varphi_p}.
\end{equation}

Reduction of the product of two rotation matrices:
\begin{multline}
\label{eq:DDD1A}
\mathcal{D}^{\ell_1\ast}_{m_1^\prime,m_1} \mathcal{D}^{\ell_2}_{m_2^\prime,m_2}=\\ (-1)^{m_1^\prime-m_1}\sum_{JM_JM^\prime_J}(-1)^{M_J-M^\prime_J}  (J) \left( \begin{array}{ccc} \ell_1 & \ell_2 &J \\-m_1^\prime & m_2^\prime& -M^\prime_J\end{array}  \right) \left( \begin{array}{ccc} \ell_1 & \ell_2 &J \\-m_1& m_2& -M_J\end{array}  \right)\mathcal{D}^{J}_{M^\prime_J,M_J}
\end{multline}
\begin{multline}
\label{eq:DDD2A}
\mathcal{D}^1_{k_1^\prime,\pm 1} \mathcal{D}^1_{k_2^\prime,\pm1}=\sum_{TM_TM^\prime_T}(-1)^{M_T-M^\prime_T}  (T)  \left( \begin{array}{ccc} 1 & 1 &T \\ k_1^\prime& k_2^\prime& -M^\prime_T\end{array}  \right) \left( \begin{array}{ccc} 1 & 1 &T \\ \pm1 & \pm1  & -M_T\end{array}  \right)\mathcal{D}^{T}_{M^\prime_T,M_T}
\end{multline}

Summation over indices $m_1$ and $m_2$:
\begin{multline}
\label{eq:DDD3A}
\sum_{m_1 m_2}(-1)^{m_2 -m_1} \left( \begin{array}{ccc} \ell_1 & \ell_2 &2 \\m_1 & -m_2& +1\end{array}  \right)  \left( \begin{array}{ccc} \ell_1 & \ell_2 &J \\-m_1& m_2& -M_J\end{array}  \right)  = (-1)^{M_J}(-1)^{\ell_1+ \ell_2 +J }\frac{\delta_{2,J} \delta_{1,M_J}}{(J)}.
\end{multline}

Averaging of the product of three rotation matrices over all molecular orientation angles:
\begin{equation}
\label{eq:3rotmatA}
\frac{1}{8\pi^2} \int d^3(\alpha\beta\gamma)\, \mathcal{D}^{2}_{M_J^\prime,+1} \mathcal{D}^{1}_{-k^{\prime\prime},\mp1}  \mathcal{D}^{T}_{M^\prime_T,M_T}   =  \left( \begin{array}{ccc} 2& 1&T\\ M_J^\prime  & -k^{\prime\prime}  & M^\prime_T \end{array}  \right)\left( \begin{array}{ccc} 2& 1&T\\ +1  & \mp1& M_T\end{array}  \right).
\end{equation}

\end{document}